\documentclass{optica-article}

\journal{oe}

\articletype{Research Article}

\begin{document}

\title{Properties of Two-Mode Quadrature Squeezing from Four-wave Mixing in Rubidium Vapor}
\author{Lu\'{i}s E.~E.~de Araujo,\authormark{1,2} Zhifan Zhou,\authormark{1} Matt DiMario, \authormark{1} B. E. Anderson, \authormark{3} Jie Zhao, \authormark{1} Kevin M. Jones, \authormark{4} and Paul D. Lett\authormark{1,5}}

\address{{$^1$}Joint Quantum Institute, National Institute of Standards and Technology and the University of Maryland, College Park, Maryland 20742, USA\\
\authormark{2}Institute of Physics Gleb Wataghin, University of Campinas (UNICAMP), 13083-859 Campinas, S\~{a}o Paulo, Brazil\\
\authormark{3}Department of Physics, American University, Washington DC 20016, USA\\
\authormark{4}Department of Physics, Williams College, Williamstown, Massachusetts 01267, USA\\
\authormark{5}Quantum Measurement Division, National Institute of Standards and Technology, Gaithersburg, Maryland 20899, USA}

\email{\authormark{*}lett@umd.edu} 



\begin{abstract}
We present a study of homodyne measurements of two-mode, vacuum-seeded, quadrature-squeezed light generated by four-wave mixing in warm rubidium vapor. Our results reveal that the vacuum squeezing can extend down to measurement frequencies of less than 1 Hz, and the squeezing bandwidth, similar to the seeded intensity-difference squeezing measured in this system, reaches up to approximately 20 MHz for typical pump parameters. By dividing the squeezing bandwidth into smaller frequency bins, we show that different sideband frequencies represent independent sources of two-mode squeezing. Such frequency bins may provide useful qumodes for quantum information processing experiments. We also investigate the impact of group velocity delays on the correlations in the system.
\end{abstract}

\section{Introduction}

Four-wave mixing\,(4WM) in warm Rb vapor has been successfully used to generate two-mode squeezed light for a number of years~\cite{boyer2008, clark2012imaging}. Seeded, bright-beam two-mode squeezing has been used in many quantum sensor studies~\cite{pooser2015MEMS, Wu2019Membrane, lawrie2019Review, ather2023quantum}, as it is easy to directly detect the amplitude difference squeezing on a balanced photodetector~\cite{mccormick2007strong}. While it has been studied, relatively less work has been done using homodyne detection techniques due to the extra complexity required in generating local oscillators, as well as the additional noise coupled into this system by the local oscillators, as discussed below. Intensity-difference measurements, however, do not provide access to the entanglement of the amplitude and phase quadratures necessary for many quantum information processing tasks. Moreover, homodyne detection is essential for detecting the low light levels of most vacuum-squeezed light~\cite{boyer2008}. 

\begin{figure}[htbp]
\centering
\includegraphics[width=\linewidth]{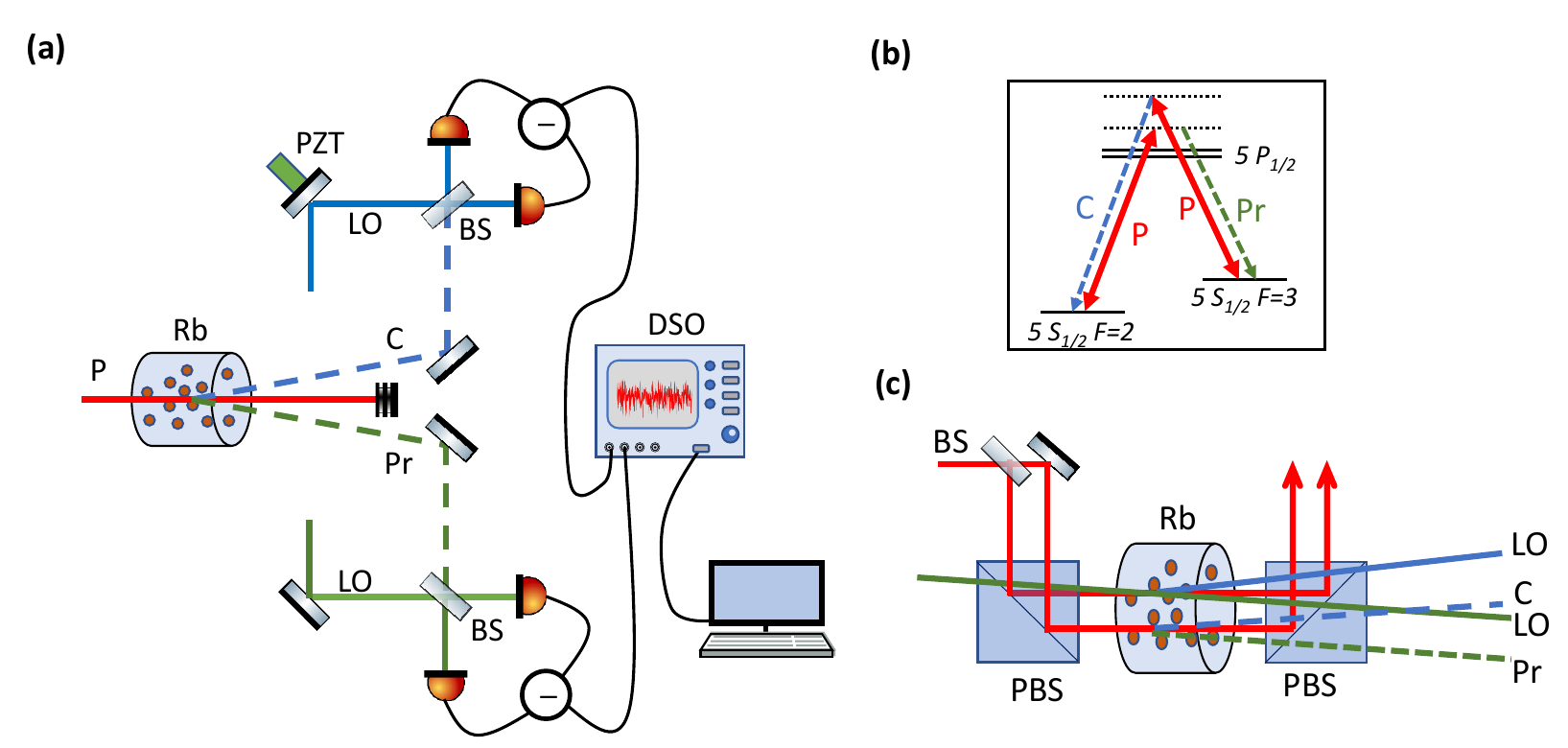}
\caption{(a) Experimental setup and (b) energy level diagram of the four-wave mixing (4WM) process in $^{85}$Rb used to generate and measure two-mode squeezed light. The cell is heated to a temperature of $\approx120~^{\circ}$C to achieve the required vapor density. The pump beam (P) generates twin probe (Pr) and conjugate beams (C) in two-mode squeezed states. Non-polarizing 50/50 beamsplitters (BS), a piezoelectric transducer (PZT), local oscillator (LO) fields, and a 1~GHz digital sampling oscilloscope (DSO) are used for the homodyne detection schemes. (c) The pump beam is split into equal parts that are used to generate the local oscillators and signal beams independently in the same cell.}
\label{fig1}
\end{figure}
 
In this paper, we present homodyne measurements of vacuum-seeded quadrature squeezing, demonstrating that the particular 4WM process sketched in Fig.1 is a useful system for quantum sensing and continuous-variable quantum information investigations.  In particular we show that it can generate squeezing down to measurement frequencies below 1 Hz (we measure more than -4 dB here). These results are obtained without using cavity-based field-enhancement or any feedback or active stabilization, just single-pass gain. 
This is achieved in spite of the fact that measuring multi-spatial-mode two-mode squeezing from a free-space system such as this, where the modes are not defined by a cavity, presents extra challenges. Specifically, the local oscillator (LO) overlap with the desired modes must be much higher than that required for measuring single-mode squeezing from a cavity-defined optical mode~\cite{gupta2020effect}. 
This is because the noise penalty associated with collecting light from the surrounding modes is significantly higher. These uncorrelated squeezed states are extremely noisy, as opposed to the vacuum modes that typically surround a cavity transmission mode.

Intensity-difference squeezed light from this 4WM system has been used to improve the sensitivity of some sensor applications~\cite{pooser2015MEMS, Wu2019Membrane, lawrie2019Review, ather2023quantum}, but entangled quantum sensors based on either frequency or spatial modes may also require homodyne measurements~\cite{braunstein2005, AndersonPRA, AndersonOptica, Furusawa2013, Furusawa2019, Anderson2019, guo2020distributed, zhu2021}, and quantum information processing applications certainly will. 
For this purpose we demonstrate that the bandwidth of the two-mode squeezing in this system can be divided into independent frequency bins, or continuous-variable qumodes, in analogy to discrete qubits. The overall bandwidth of squeezing in this Rb-based 4WM system is relatively small (approximately 20 MHz), making it perhaps impractical for scaling up for quantum computing applications. On the other hand, the limited bandwidth allows for the digitization of the measurements across the entire bandwidth at once. This sort of measurement of many qumodes simultaneously is proposed in~\cite{ferrini2013compact, Cai2015, ferrini2016direct} as a "direct approach" to Gaussian measurement-based quantum computation. In such experiments a series of Gaussian operations are rotated into the same quadrature before measurement and can be compressed to a quantum operation depth of one (all measurements can be made simultaneously). We show the independence of the frequency qumodes in this system, and their suitability for such experiments.

Finally, we also demonstrate the effect on the squeezing spectrum of a delay between the spatially-separated twin beams.  While the effect itself is already well-understood, its importance for such continuous-variable applications is made clear in this context.
Thus, we show that 4WM is an interesting system that generates strong quantum correlations across multiple frequencies, as well as spatial modes, and lends itself to simple demonstration experiments in quantum information processing.  

\section{Experiment}

Figure 1 shows a schematic of the experimental system. A pump beam with a power of approximately 800 mW is divided into two using a 50/50 beamsplitter, and both beams are directed in parallel into a 12 mm long Rb vapor cell maintained at $\approx120^\circ$C. These pump beams with $\approx$ 650 $\mu$m beam diameter produce the signals and local oscillators separately in two pumped regions of the vapor, as described in~\cite{boyer2008}. The generation of the local oscillators is shown in Fig.~1(c). A seed beam for the local oscillators is produced by double-passing a small fraction of the pump beam through a 1.5 GHz acousto-optical modulator, which downshifts the frequency of the light by $\approx$ 3 GHz (the ground-state splitting in $^{85}$Rb). The seed and pump beams are crossed at a small angle of approximately 0.3 degrees to generate local oscillators with a power of about 0.5 mW. These local oscillators are bright twin beams themselves at symmetric probe and conjugate frequencies with respect to the pump, and are usable as phase references at each frequency. Although individually these beams are noisy, displaced thermal beams, they do not compromise the detection sensitivity when used as local oscillators in a balanced homodyne detector as long as the beamsplitter is carefully balanced and the  detector's common mode noise suppression is sufficiently high.

The probe and conjugate beams, which constitute the two-mode squeezed vacuum signals, are aligned with the local oscillators and detectors. It is done by first seeding the 4WM process in the same way as for the local oscillators and aligning the optics for mode matching with visibilities higher than 98 \%. That seed is then blocked to allow for vacuum-seeded measurements. The relative phase of the LOs is set using a noise-locking technique~\cite{mckenzie2005}, in which the phase of one LO beam is adjusted so that the dual homodyne measurement signals stay stabilized for measuring the corresponding squeezed quadrature. We can lock on either the amplitude difference squeezing signal or the phase sum squeezing signal. The resulting signals of the probe and conjugate from the two balanced homodyne detectors are digitized and recorded by an 8-bit digital sampling oscilloscope. The shot-noise level is obtained by Fourier transforming the autocorrelation of the difference of the two signals from vacuum-seeded (input signals blocked) homodyne measurements. The two-mode squeezing signal is obtained from the difference (or sum) of the two signals with squeezed vacuum injected into the homodyne detectors.  


\begin{figure}[htbp]
\centering
\includegraphics[width=12cm]{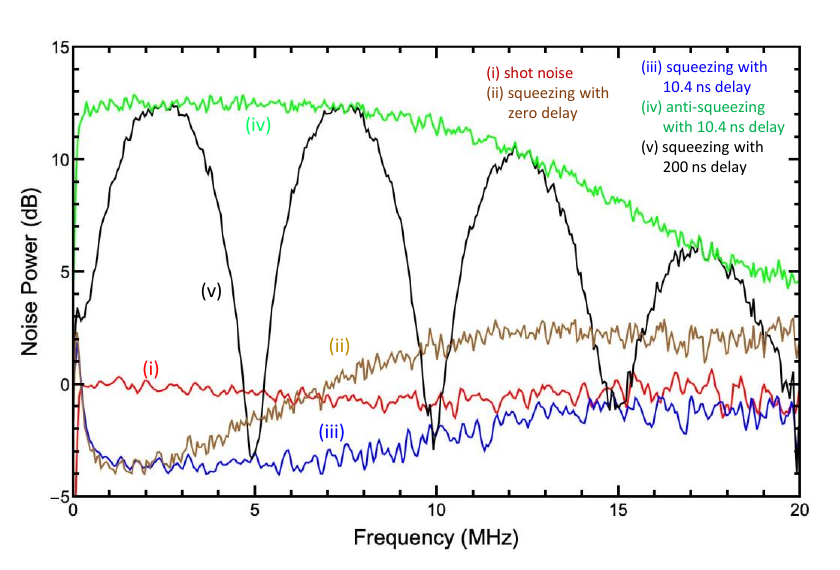}
\caption{The two-mode squeezed vacuum spectra and the effect of group velocity delay between probe and conjugate. The traces show the squeezing spectrum with (i) shot noise, where the vacuum noise is measured by the homodyne measurements, (ii) the default squeezing (zero additional delay in conjugate signal) showing the delay from the group velocity difference between the probe and conjugate in the Rb vapor, (iii) squeezing with an optimum delay of 10.4 ns in the conjugate signal to compensate for the group velocity delay difference from the 4WM, (iv) anti-squeezing with the optimum delay of 10.4 ns, and (v) squeezing with a large delay of 200\,ns. The plot clearly highlights the importance of selecting the optimum delay for achieving the maximum squeezing bandwidth. The intentional long delay is introduced to show the impact of delay on the acquired squeezing. All of the plots have the electronic noise subtracted. } 
\label{fig2}
\end{figure}

Figure 2 shows the measured two-mode squeezing and shot noise signals. Curve (i) shows the shot-noise level, and curve (ii) shows the measured squeezing level with the default delay setting (zero additional delay in the conjugate signal). The measurement is influenced by the delay coming from the group velocity difference between the probe and conjugate beams in the 4WM process. We consider the effect of the group velocity delay on the intensity-difference squeezing signal, which has been discussed in previous 4WM work~\cite{boyer2007,boyer2008} and is consistent with the discussions in~\cite{machida1989observation}. Because the twin beams in our system are separated in frequency by approximately 6\,GHz they experience different group velocities in the Rb vapor at low intensities. The group velocities vary with pump detuning and the intensities of the various beams. At higher probe and conjugate intensities, the group velocities lock together at a fixed delay, but the vacuum-squeezed beams also experience this difference, leading to an approximately 10\,ns arrival time difference between the correlated probe and conjugate signals for typical pump conditions~\cite{boyer2007}. As the dispersion delays light at the probe frequency more, an additional BNC cable can be added to the conjugate homodyne measurement to compensate for the delay. Although the local oscillators\,(LOs) also experience this differential delay, it can be ignored since they are treated as classical phase references. 

In curve (iii) the optimal squeezing spectrum is obtained by Fourier transforming the autocorrelation of the amplitude-difference signals at an optimum delay of approximately 10.4\,ns where the squeezing is maximized, and it smoothly goes from maximum squeezing at low frequencies to the shot noise level around the 20\,MHz bandwidth of the 4WM process. In curve (iv) the anti-squeezing spectrum is obtained by Fourier transforming the autocorrelation of the amplitude-sum of the two signals, again at the optimum delay of approximately 10.4 ns. The anti-squeezing level shows the excess noise imposed on the signal over the 4WM process bandwidth, and while the squeezing level can easily be affected by small inefficiencies or added noise, the anti-squeezing noise levels are hardly altered by these effects. In the ideal case of pure squeezing and no added noise the squeezing should go as far below the shot noise level as the anti-squeezing goes above this level on a logarithmic scale. We find that the anti-squeezing levels are larger than the measured squeezing levels due to difficulties in measuring uncontaminated squeezing, as discussed above and in Ref.~\cite{gupta2020effect}. In curve (v) an intentional long delay between the signals (200\,ns) is introduced in software to demonstrate the impact of the group velocity delay on the acquired squeezing, which leads to an apparent oscillation in the squeezing spectrum at $1/T$ (5MHz). This curve oscillates as cos($\omega T$) between the maximum squeezing and the maximum anti-squeezing levels that can be measured, where $\omega$ is the measurement frequency and $T$ is the delay between the beams.  It is clear that a large intentional delay between the twin beams (introduced digitally here) can be used to locate the envelope of both the maximum squeezing and anti-squeezing in the system, and helps to define the optimum delay required to produce the best squeezing spectrum.


\begin{figure}[htbp]
\centering
\includegraphics[width=\linewidth]{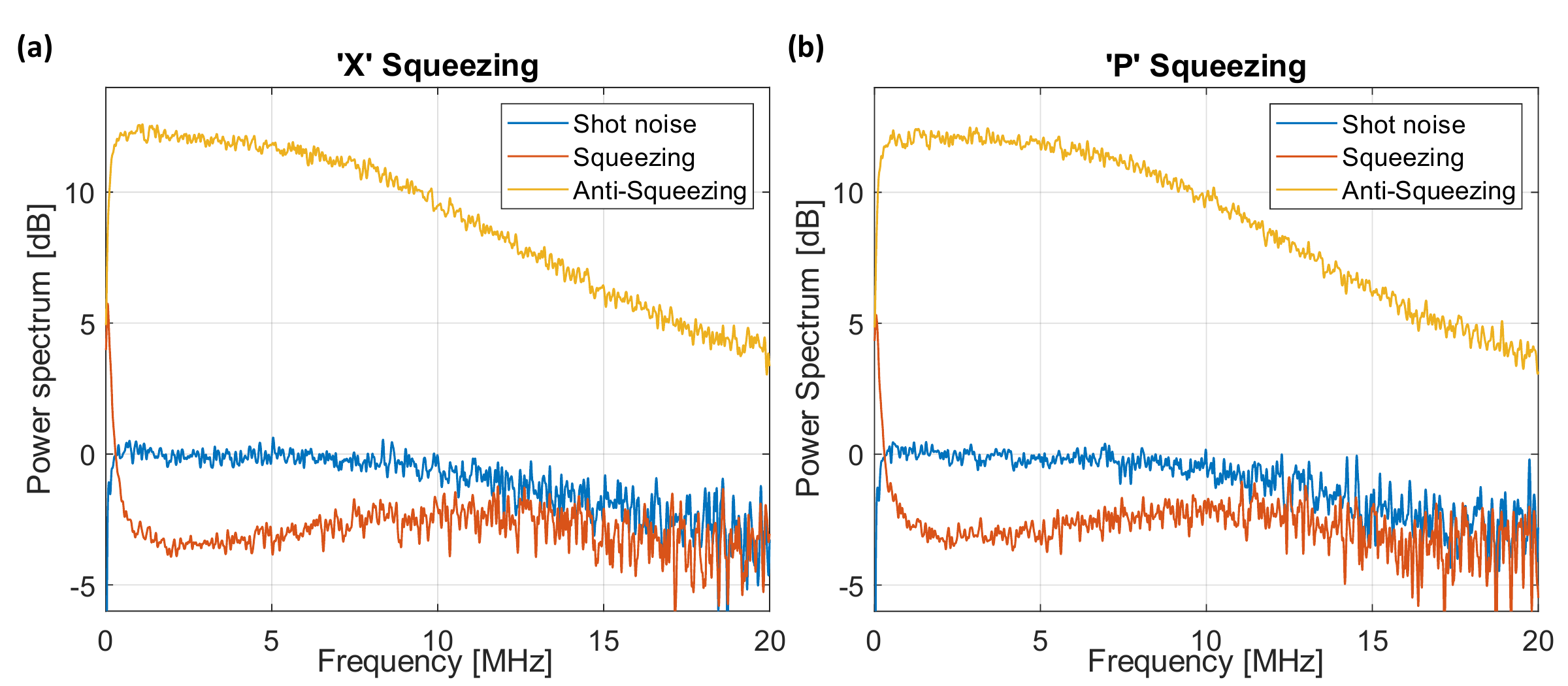}
\caption{Plot of squeezing and anti-squeezing levels for the orthogonal quadratures of the two-mode squeezed state generated by four-wave mixing in Rb vapor as a function of the measurement frequency. (a) The "X," and (b) the "P" squeezing represent the amplitude-difference and phase-sum squeezing between the probe and conjugate signals. In the experiment, depending on whether the difference or the sum of the two homodyne measurements is injected into the locking circuit, the relative phase is stabilized for measuring the squeezing or the anti-squeezing signal. 
The blue curve is the shot noise level (homodyning a vacuum input). The red and yellow curves are the squeezing and anti-squeezing levels, respectively, with a 10.4\,ns delay in the conjugate signal. The spectra all have electronic noise subtracted, resulting in the noise above 10 MHz, as the electronic noise approaches the squeezing and shot noise signals in that range. The squeezing displayed here does not extend closer to DC due to the frequency response of an RF power splitter used to enable locking in the measurement circuit.} 
\label{fig3}
\end{figure}

\begin{figure}[htbp]
\centering
\includegraphics[width=12cm]{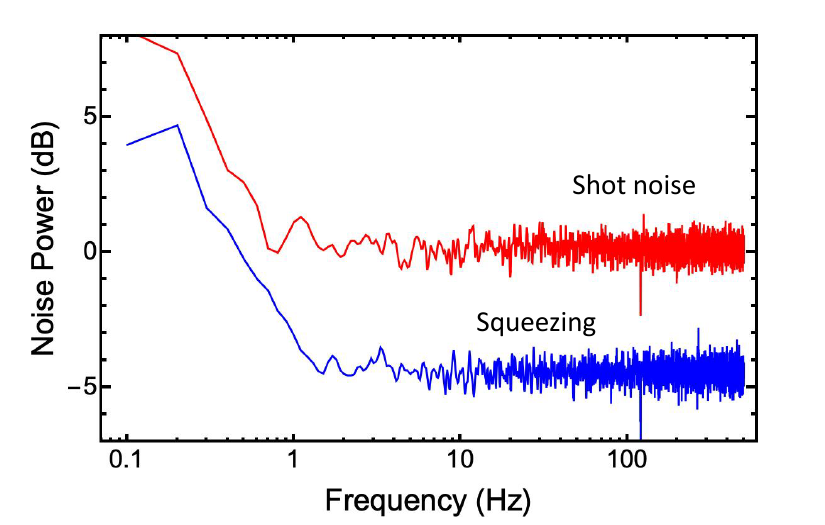}
\caption{Homodyne measurement of vacuum quadrature squeezing at low frequencies. The blue trace shows the homodyne power spectrum with a two-mode squeezed vacuum state present, and the red trace shows the spectrum with a vacuum state present, which is the shot noise level. The vertical axis shows the noise power spectral density in units of dB relative to the shot noise level. The horizontal axis shows the Fourier measurement frequency in Hz. The data were taken over a 10-second interval. The RF power splitters in the measurement circuits are replaced with BNC tees to avoid the low-frequency cut-off.}
\label{fig4}
\end{figure}

The squeezing levels for the amplitude and phase quadratures are entirely symmetric in this system, as there are no cavity instabilities, for example, to introduce excess phase noise. The "X" and "P" squeezing represents the amplitude difference and phase sum squeezing between the probe and conjugate. In the experiment, depending on whether the difference signal or the sum signal of the two homodyne measurements is injected into the locking circuit, the relative phase is stabilized for measuring the squeezing or the anti-squeezing signal. As seen in Fig.\,3, the bandwidth of the 4WM process is apparent in the width of the excess noise resulting from the 4WM gain in the anti-squeezing spectra. The squeezing spectra, when adjusted for group-velocity delays, exhibit a gradual increase in noise with frequency, eventually reaching shot noise at around 15\,MHz, where the 4WM gain becomes insignificant.



The low-frequency range for bright-beam-seeded intensity-difference squeezing is typically limited by the seed noise and the low-frequency power imbalance in the beams. The low-frequency squeezing performance can be recovered by a dual-seeding technique that balances the powers of the beams~\cite{wu2019}.  With the vacuum-seeded process there is no imbalance and this is not necessary. In typical cavity-based optical parametric oscillator (OPO) systems used to generate squeezing, the coupling of acoustic noise from the cavity can also limit the performance at low frequencies.  This has led to the construction of monolithic cavities where the mirrors are made directly on the surface of a nonlinear crystal to reduce this noise. In a single-pass-gain system like ours, we are free of any coupling to such cavity acoustic noise. The 4WM process will, in principle, generate squeezing down to DC sideband frequencies. Fig. 4 shows that sub-1 Hz 2-mode squeezing can be measured in a 10 s measurement period. At some point, instabilities in the beam pointing, coupled to detector nonuniformity, become a problem in obtaining low-frequency squeezing close to  DC~\cite{stefszky2012balanced}. In Figs. 2 and 3, a low-frequency roll-off of the squeezing signal around 300 kHz is observed due to the low-frequency cut-off from the  RF power splitters that are inserted to split part of the signal to the lock circuit, which are avoided in Fig. 4. by replacing them with BNC tees.  

\begin{figure}[htbp]
\centering
\includegraphics[width=13.5cm]{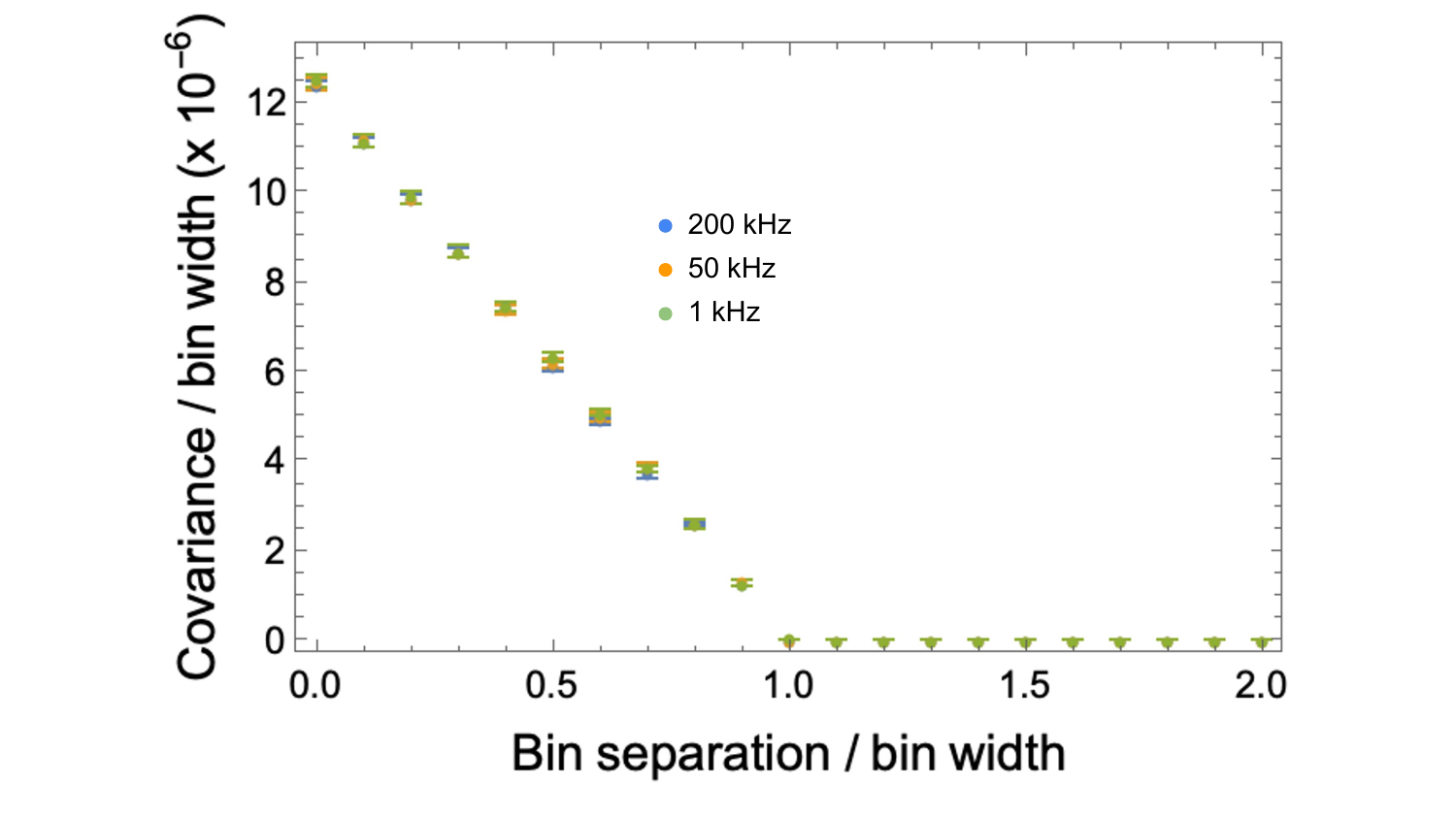}
\caption{The noise covariance of equal-sized probe and conjugate frequency bins plotted versus the shift of the center frequency of one of the bins.  The probe bin is centered at 1 MHz and the conjugate is shifted to higher frequencies.  The spacing or shift of the conjugate bin is given in units of the bin width for bin sizes of 200 kHz, 50 kHz, and 1~kHz.  The noise covariance is normalized to the bin width.  A square filter in frequency is used. The error bars represent one standard deviation statistical uncertainties.}
\label{fig5}
\end{figure}

Bright-beam intensity-difference squeezing down to -9 dB~\cite{Quentin2011} has been reported in this 4WM system, where the entirety of the correlated bright beams can be collected. Any additional uncorrelated modes that are collected will be vacuum-seeded and will contribute relatively little to the measured signals in comparison to the bright beams. In making homodyne measurements of vacuum-seeded twin beams, however, the additional uncorrelated modes that might be collected are also vacuum-seeded squeezed light and contribute a large excess noise to the signal. Such mode-matching issues have typically limited our homodyne squeezing measurements to -5 dB or less~\cite{gupta2020effect}. To increase the amount of measured squeezing, one could try to define the desired modes better using spatial filters. Proper alignment could replace the surrounding uncorrelated vacuum-squeezed spatial modes that are partly picked up with much less-noisy unsqueezed vacuum modes.


The squeezing in this system can be viewed as a continuous spectrum of sideband frequencies up to $\approx$20 MHz or, alternatively, as a series of independent two-mode squeezers at these sideband frequencies. To explore this, we consider frequency bins with certain bin sizes from the probe and conjugate signals, which are obtained by numerically filtering and processing the time-series data. The filtering here is a square filter and is discussed further below. The noise in the bins is used to evaluate the correlation behavior via the covariance. Figure 5 shows the covariance of the probe and conjugate frequency bins as a function of their frequency-space overlap. The spacing between the center of the bins is plotted on the x-axis, and the covariance is on the y-axis. The data are shown for bin widths of 200 kHz, 50 kHz, and 1 kHz. The bin from the probe is centered at 1 MHz in the measured spectrum, and the conjugate bin, which has the same size, is progressively shifted to higher frequencies to change the overlap. For a bin spacing of zero, probe and conjugate bins are fully overlapped, and this frequency-space overlap decreases as the center-spacing increases. For a bin spacing greater than one (where the bins are adjacent to each other), probe and conjugate bins no longer overlap as their spacing is greater than their widths. The figure shows that down to a frequency-bin size of 1kHz, frequency bins with non-overlapping offsets from the probe and conjugate local oscillators act as independent sources of 2-mode vacuum squeezing. The ability to digitally define frequency bins in this way allows for manipulation of the squeezing spectrum and the use of these bins as qumodes in quantum state processing. 

The frequency bins considered in this study were obtained by numerically filtering and processing the time-series data. Several different filter functions were tried, yielding substantially the same results. The plot shown used a hard-edged "square" filter, which is non-causal, but allows the edges of the bins to be well-defined. Using a realistic causal filter does not change the results beyond requiring the frequency bins to be separated by some amount beyond the nominal frequency bin edges to account for the sloping sides of the filter function.  The bin width can be made even smaller, and we have verified the independence of bins down to bin widths of 5 Hz.

\section{Conclusion}


In summary, the Rb-vapor-based 4WM system provides a valuable platform for generating and studying homodyne-detected squeezed light. Although homodyne detection of two-mode quadrature squeezing can be more complicated than the direct detection of bright-beam intensity-difference squeezing, it is necessary to detect the entanglement between the beams that can be used for quantum information processing and sensing applications. The squeezing bandwidth of the 4WM  system used here is limited to about 20 MHz. Although this small bandwidth limits the number of modes and potential measurement speed in such a system, an advantage is the ability to digitize the entire spectrum at once and to select portions of the squeezing spectrum and frequency modes in software.  This makes it a good system for demonstration experiments.

This system holds promise for further work in quantum information processing and quantum technology. This could include exploring the potential of the system for cluster state generation and quantum state processing~\cite{ferrini2013compact, Cai2015, ferrini2016direct}. Progress can also be expected in techniques to increase the amount of measured squeezing and to reduce further the noise introduced in the detection. In addition, we have shown that squeezing signals can be obtained below 1 Hz in measurement frequency in this system without requiring feedback or laser frequency stabilization.





\section*{Funding}
Air Force Office of Scientific Research (FA9550- 16-1-0423). L.~E.~E.~de Araujo acknowledges the financial support of grant \#2019/24743-9, S\~{a}o Paulo Research Foundation (FAPESP). This research was performed while
Matthew DiMario held a National Research Council Research Associateship at NIST. 

\section*{Acknowledgments}
We acknowledge Alessandro Restelli for the help with the lock circuits. 

\section*{Disclosures} 
The authors declare no conflicts of interest.

\bibliography{sample}

\begin{thebibliography}{10}
\newcommand{\enquote}[1]{``#1''}

\bibitem{boyer2008}
V.~Boyer, A.~M. Marino, R.~C. Pooser, and P.~D. Lett, \enquote{Entangled images
  from four-wave mixing,} {\protect\JournalTitle{Science}} \textbf{321}, 544
  (2008).

\bibitem{clark2012imaging}
J.~B. Clark, Z.~Zhou, Q.~Glorieux, A.~M. Marino, and P.~D. Lett,
  \enquote{Imaging using quantum noise properties of light,}
  {\protect\JournalTitle{Optics Express}} \textbf{20}, 17050--17058 (2012).

\bibitem{pooser2015MEMS}
R.~C. Pooser and B.~Lawrie, \enquote{Ultrasensitive measurement of
  microcantilever displacement below the shot-noise limit,}
  {\protect\JournalTitle{Optica}} \textbf{2}, 393--399 (2015).

\bibitem{Wu2019Membrane}
X.~Wei, J.~Sheng, Y.~Wu, W.~Liu, and H.~Wu, \enquote{Twin-beam-enhanced
  displacement measurement of a membrane in a cavity,}
  {\protect\JournalTitle{Applied Physics Letters}} \textbf{115}, 251105 (2019).

\bibitem{lawrie2019Review}
B.~J. Lawrie, P.~D. Lett, A.~M. Marino, and R.~C. Pooser, \enquote{Quantum
  sensing with squeezed light,} {\protect\JournalTitle{{ACS} Photonics}}
  \textbf{6}, 1307--1318 (2019).

\bibitem{ather2023quantum}
H.~Ather, H.~An, H.~Owens, S.~Alajlouni, A.~Shakouri, and M.~Hosseini,
  \enquote{Quantum sensing of thermoreflectivity in electronics,}
  {\protect\JournalTitle{Phys. Rev. Appl.}} \textbf{19}, 044040 (2023).

\bibitem{mccormick2007strong}
C.~McCormick, V.~Boyer, E.~Arimondo, and P.~Lett, \enquote{Strong relative
  intensity squeezing by four-wave mixing in rubidium vapor,}
  {\protect\JournalTitle{Optics Letters}} \textbf{32}, 178--180 (2007).

\bibitem{gupta2020effect}
P.~Gupta, R.~W. Speirs, K.~M. Jones, and P.~D. Lett, \enquote{Effect of
  imperfect homodyne visibility on multi-spatial-mode two-mode squeezing
  measurements,} {\protect\JournalTitle{Optics Express}} \textbf{28}, 652--664
  (2020).

\bibitem{braunstein2005}
S.~L. Braunstein and P.~van Loock, \enquote{Quantum information with continuous
  variables,} {\protect\JournalTitle{Rev. Mod. Phys.}} \textbf{77}, 513 (2005).

\bibitem{AndersonPRA}
B.~E. Anderson, B.~L. Schmittberger, P.~Gupta, K.~M. Jones, and P.~D. Lett,
  \enquote{Optimal phase measurements with bright- and vacuum-seeded {SU}(1,1)
  interferometers,} {\protect\JournalTitle{Phys. Rev. A}} \textbf{95}, 063843
  (2017).

\bibitem{AndersonOptica}
B.~E. Anderson, P.~Gupta, B.~L. Schmittberger, T.~Horrom, C.~Hermann-Avigliano,
  K.~M. Jones, and P.~D. Lett, \enquote{Phase sensing beyond the standard
  quantum limit with a variation on the {SU}(1, 1) interferometer,}
  {\protect\JournalTitle{Optica}} \textbf{4}, 752--756 (2017).

\bibitem{Furusawa2013}
S.~Yokoyama, R.~Ukai, S.~C. Armstrong, C.~Sornphiphatphong, T.~Kaji, S.~Suzuki,
  J.~Yoshikawa, H.~Yonezawa, N.~C. Menicucci, and A.~Furusawa,
  \enquote{Ultra-large-scale continuous-variable cluster states multiplexed in
  the time domain,} {\protect\JournalTitle{Nature Photonics}} \textbf{7},
  982--986 (2013).

\bibitem{Furusawa2019}
W.~Asavanant, Y.~Shiozawa, S.~Yokoyama, B.~Charoensombutamon, H.~Emura, R.~N.
  Alexander, S.~Takeda, J.~Yoshikawa, N.~C. Menicucci, H.~Yonezawa, and
  A.~Furusawa, \enquote{Generation of time-domain-multiplexed two-dimensional
  cluster state,} {\protect\JournalTitle{Science}} \textbf{366}, 373--376
  (2019).

\bibitem{Anderson2019}
M.~V. Larsen, X.~Guo, C.~R. Breum, J.~S. Neergaard-Nielsen, and U.~L. Andersen,
  \enquote{Deterministic generation of a two-dimensional cluster state,}
  {\protect\JournalTitle{Science}} \textbf{366}, 369--372 (2019).

\bibitem{guo2020distributed}
X.~Guo, C.~R. Breum, J.~Borregaard, S.~Izumi, M.~V. Larsen, T.~Gehring,
  M.~Christandl, J.~S. Neergaard-Nielsen, and U.~L. Andersen,
  \enquote{Distributed quantum sensing in a continuous-variable entangled
  network,} {\protect\JournalTitle{Nature Physics}} \textbf{16}, 281--284
  (2020).

\bibitem{zhu2021}
X.~Zhu, C.-H. Chang, C.~Gonź\'{a}lez-Arciniegas, A.~Pe’er, J.~Higgins, and
  O.~Pfister, \enquote{Hypercubic cluster states in the phase-modulated quantum
  optical frequency comb,} {\protect\JournalTitle{Optica}} \textbf{8}, 281
  (2021).

\bibitem{ferrini2013compact}
G.~Ferrini, J.-P. Gazeau, T.~Coudreau, C.~Fabre, and N.~Treps, \enquote{Compact
  {G}aussian quantum computation by multi-pixel homodyne detection,}
  {\protect\JournalTitle{New Journal of Physics}} \textbf{15}, 093015 (2013).

\bibitem{Cai2015}
Y.~Cai, J.~Feng, H.~Wang, G.~Ferrini, X.~Xu, J.~Jing, and N.~Treps,
  \enquote{Quantum-network generation based on four-wave mixing,}
  {\protect\JournalTitle{Phys. Rev. A}} \textbf{91}, 013843 (2015).

\bibitem{ferrini2016direct}
G.~Ferrini, J.~Roslund, F.~Arzani, C.~Fabre, and N.~Treps, \enquote{Direct
  approach to {G}aussian measurement based quantum computation,}
  {\protect\JournalTitle{Physical Review A}} \textbf{94}, 062332 (2016).

\bibitem{mckenzie2005}
K.~McKenzie, E.~E. Mikhailov, K.~Goda, P.~K. Lam, N.~Grosse, M.~B. Gray,
  N.~Mavalvala, and D.~E. McClelland, \enquote{Quantum noise locking,}
  {\protect\JournalTitle{J. Opt. B: Quantum Semiclass. Opt.}} \textbf{7}, S421
  (2005).

\bibitem{boyer2007}
V.~Boyer, C.~F. McCormick, E.~Arimondo, and P.~D. Lett, \enquote{Ultraslow
  propagation of matched pulses by four-wave mixing in an atomic vapor,}
  {\protect\JournalTitle{Phys. Rev. Lett.}} \textbf{99}, 143601 (2007).

\bibitem{machida1989observation}
S.~Machida and Y.~Yamamoto, \enquote{Observation of amplitude squeezing from
  semiconductor lasers by balanced direct detectors with a delay line,}
  {\protect\JournalTitle{Optics Letters}} \textbf{14}, 1045--1047 (1989).

\bibitem{wu2019}
M.-C. Wu, B.~L. Schmittberger, N.~R. Brewer, R.~W. Speirs, K.~M. Jones, and
  P.~D. Lett, \enquote{Twin-beam intensity-difference squeezing below 10 {H}z,}
  {\protect\JournalTitle{Optics Express}} \textbf{27}, 4769--4780 (2019).

\bibitem{stefszky2012balanced}
M.~Stefszky, C.~Mow-Lowry, S.~Chua, D.~Shaddock, B.~Buchler, H.~Vahlbruch,
  A.~Khalaidovski, R.~Schnabel, P.~K. Lam, and D.~McClelland, \enquote{Balanced
  homodyne detection of optical quantum states at audio-band frequencies and
  below,} {\protect\JournalTitle{Classical and Quantum Gravity}} \textbf{29},
  145015 (2012).

\bibitem{Quentin2011}
Q.~Glorieux, L.~Guidoni, S.~Guibal, J.-P. Likforman, and T.~Coudreau,
  \enquote{Quantum correlations by four-wave mixing in an atomic vapor in a
  nonamplifying regime: Quantum beam splitter for photons,}
  {\protect\JournalTitle{Phys. Rev. A}} \textbf{84}, 053826 (2011).

\end{thebibliography}






\end{document}